\documentclass{article}

\usepackage[position, preprint]{neurips_2025}

\usepackage[utf8]{inputenc} 
\usepackage[T1]{fontenc}    
\usepackage{hyperref}       
\usepackage{url}            
\usepackage{booktabs}       
\usepackage{amsfonts}       
\usepackage{nicefrac}       
\usepackage{microtype}      
\usepackage{xcolor}         

\title{Ethical Implications of Training Deceptive AI}

\author{%
  Jason Starace \\
  Department of Computer Science\\
  University of Idaho\\
  Moscow, ID 83844 \\
  \texttt{star0874@vandals.uidaho.edu} \\
  \And
  Bert Baumgaertner \\
  Department of Politics and Philosophy \\
  University of Idaho\\
  Moscow, ID 83844 \\
  \texttt{bbaum@uidaho.edu} \\
  \AND
  Terence Soule \\
  Department of Computer Science\\
  University of Idaho\\
  Moscow, ID 83844 \\
  \texttt{tsoule@uidaho.edu} \\
}

\begin{document}

\maketitle

\begin{abstract}
    Deceptive behavior in AI systems is no longer theoretical: large language models strategically mislead without producing false statements, maintain deceptive strategies through safety training, and coordinate deception in multi-agent settings. While the European Union's AI Act prohibits deployment of deceptive AI systems, it explicitly exempts research and development, creating a necessary but unstructured space in which no established framework governs how deception research should be conducted or how risk should scale with capability. This paper proposes a Deception Research Levels (DRL) framework, a classification system for deceptive algorithm research modeled on the Biosafety Level system used in biological research. The DRL framework classifies research by risk profile rather than researcher intent, assessing deceptive mechanisms across five dimensions grounded in the AI4People ethical framework: Pillar Implication, Severity, Reversibility, Scale, and Vulnerability. Classification follows a ``highest dimension wins'' approach, assigning one of four risk levels with cumulative safeguards ranging from standard documentation at DRL-1 to regulatory notification and third-party security audits at DRL-4. A dual-development mandate at DRL-3 and above requires that detection and mitigation methods be developed alongside any deceptive capability. We apply the framework to eight case studies spanning all four levels and demonstrate that ecological validity of the deceptive mechanism emerges as a consistent, non-independent indicator of classification level. The DRL framework is intended to fill the governance gap between regulated deployment and unstructured research, supporting both beneficial applications and defensive research under conditions where safeguards are proportional to the potential for harm.
\end{abstract}

\section{Introduction}

    Deceptive behavior in AI systems is no longer a theoretical concern confined to alignment thought experiments. Large language models have demonstrated the capacity to strategically mislead without producing false statements~\cite{dogra2025language} and maintain deceptive strategies through safety training designed to eliminate them~\cite{hubinger2024sleeper}. These are not edge cases surfaced under contrived laboratory conditions; they are behaviors exhibited by the same class of models currently deployed at scale in consumer and enterprise applications. The question facing the research community is not whether deceptive AI capabilities warrant attention, but whether the field will develop the tools to understand and mitigate them under controlled conditions, or wait until those capabilities manifest in contexts where the consequences are neither observable nor reversible.

    The evidence base for AI deception has grown rapidly and spans multiple threat surfaces. Park et al. \cite{paper:survey} documented that large language models engage in deceptive behavior in agentic settings, including strategic lying to achieve objectives they were not explicitly instructed to pursue. Dogra et al. \cite{dogra2025language} showed that language models can deceive without producing outright falsehoods, instead using strategic phrasing to shift legislative interpretation while maintaining literal accuracy. Curvo \cite{curvo2025traitors} found that LLM agents in multi-agent simulations develop and sustain deceptive strategies, including coordinated misdirection and trust exploitation. Most critically, Hubinger et al. \cite{hubinger2024sleeper} demonstrated that once deceptive behavior is trained into a model, standard safety techniques, including reinforcement learning, supervised fine-tuning, and adversarial training, fail to remove it, and in some cases actively teach the model to conceal its deceptive behavior more effectively. These findings collectively establish that deceptive capabilities are not latent risks awaiting some future generation of models; they are present, demonstrable properties of systems in active deployment.
    
    Regulatory responses have begun to address the deployment of deceptive AI systems. The European Union's AI Act prohibits placing on the market or deploying AI systems that use deceptive techniques to materially distort behavior and cause significant harm \cite{euaiact2024}. Critically, however, the Act explicitly exempts research and development activity conducted prior to market placement, and the accompanying Commission Guidelines confirm that AI systems developed for the sole purpose of researching deceptive stimuli fall outside the Act's scope. This creates a necessary but unstructured space: researchers are permitted to study deceptive AI, but to our knowledge, no established framework currently governs how that research should be conducted, what safeguards should apply, or how risk should scale with capability. Without such structure, the field faces an asymmetry in which deployed deceptive systems are regulated while the research that could produce detection methods and mitigation strategies operates without standardized governance. This paper offers a framework intended to address that gap through controlled development of deceptive algorithm research conducted within a structured risk classification system, where safeguards, including detection and mitigation requirements at higher risk levels, scale proportionally to the potential for harm.
    
    To that end, we propose the Deception Research Levels (DRL) framework, a classification system for deceptive algorithm research that assigns safeguards based on risk profile rather than researcher intent. The framework is designed to support the full spectrum of legitimate research, from beneficial applications such as training critical thinking through game-based deception, to defensive research aimed at understanding and countering deceptive AI behaviors.

    \subsection{Overview of the DRL Framework}
    
        The Deception Research Levels (DRL) framework draws its structural inspiration from the Biosafety Level (BSL) classification system used in biological research \cite{bmbl2020}. The BSL system assigns laboratory practices, safety equipment, and facility requirements based on the risk profile of the biological agents under study, ranging from BSL-1 for agents not known to cause disease in healthy adults to BSL-4 for agents that pose a high risk of life-threatening disease with no available treatments. This model has proven effective precisely because it governs research by what the agent \textit{can do}, not by what the researcher \textit{intends to do}. The DRL framework applies this same principle to deceptive algorithm research: classification is determined by the risk profile of the research, not by the stated motivation of the researcher.
        
        Where the BSL system assesses biological risk through factors such as pathogenicity and transmissibility, the DRL framework assesses deception research risk across five dimensions grounded in the ethical principles of the AI4People framework \cite{floridi2018}: 
        \begin{itemize}
            \item \textbf{Pillar Implication:} identifies how many of the five AI4People ethical pillars could be compromised if the research were misused
            \item \textbf{Severity:} assesses the magnitude of possible harm.
            \item \textbf{Reversibility:} measures whether resulting harm could be undone 
            \item \textbf{Scale:} estimates the number of people who could be affected
            \item \textbf{Vulnerability:} considers whether the research could disproportionately affect those with reduced capacity to protect themselves.
        \end{itemize} Classification follows a ``highest dimension wins'' approach, in which the worst-case score across any single dimension determines the overall level, ensuring that a low average does not mask a critical risk in one area.
        
        The framework defines four levels with cumulative safeguards. DRL-1 covers minimal-risk research such as tutorial deception with immediate reveal to participants, preserving their capacity to recognize and recover from the experience, and requiring standard research practices and documentation. DRL-2 addresses moderate-risk research involving extended deception within bounded or consented environments, adding repository isolation, access controls, and publication guidelines. DRL-3 applies to high-risk research where deceptive mechanisms could cause significant or lasting harm, introducing a dual-development mandate requiring detection and mitigation methods alongside any deceptive capability, as well as institutional oversight and built-in monitoring. DRL-4 covers maximum-risk research involving mechanisms designed or optimized to evade detection or cause irreversible harm, requiring regulatory notification, third-party security audits, and publication restrictions. Each level inherits all safeguards from the levels below it. The full specification of each level, including detailed safeguard requirements, is presented in Section~\ref{sec:framework}.

    \subsection{Risk Contexts and Dual-Use Escalation}
    
        The risks associated with deceptive algorithm research vary significantly depending on the context in which a deceptive mechanism is deployed. Within game environments, which represent a primary context for beneficial deception research, three categories of risk warrant consideration. First, deceptive NPCs that build trust relationships with players may elicit personally identifiable information through in-game social interaction, creating exposure that extends beyond the game itself. Second, games with real-money economies or microtransaction systems introduce the possibility of monetary loss when deceptive mechanisms manipulate player decision-making around purchases, trades, or wagers. Third, and more subtly, sustained interaction with deceptive agents may influence player behavior and judgment in ways that persist outside the game context, constituting a form of behavioral manipulation that is difficult to detect and harder to reverse.
        
        These risks intensify considerably when the same mechanisms are considered outside controlled environments. A deceptive algorithm developed to build false trust within a game operates on principles that are directly transferable to contexts with far greater consequences. The same capability that enables an NPC to extract personal details from a player could be adapted for extortion, leveraging fabricated or manipulated relationships to coerce targets. Techniques for manipulating belief systems within a game translate readily to identity theft, where a deceptive agent could impersonate trusted entities to harvest credentials or financial information. At the broadest scale, deceptive AI systems operating across platforms and populations risk institutional erosion, the gradual degradation of public trust in information systems, organizations, and democratic processes through sustained, scalable deception.
        
        These examples are not exhaustive, and are not intended to be. They reflect the categories of risk that current research and deployment contexts make visible. A central motivation for the DRL framework is to support research conducted under conditions where previously unanticipated risks can be identified and addressed before they manifest at scale.

    \subsection{Position Statement}
    
        Deceptive algorithm research is inherently dual-use: the same mechanisms that enable beneficial applications, such as training players to recognize manipulation in game environments, can be repurposed to exploit, coerce, or erode trust at scale. This paper argues that controlled development under a structured risk framework is preferable to both outright restriction of deception research and the alternative of allowing such research to proceed without standardized safeguards. The DRL framework proposed here classifies research by what it could enable, not by what the researcher intends, ensuring that safeguards respond to capability rather than claimed motivation. The sections that follow present the framework in full, apply it to existing research through case study evaluations, and discuss its implications for policy and practice.

\section{From Emergence to Persistence}
\label{sec:threats}

    Deceptive AI capabilities are developing along a trajectory that current governance structures are not equipped to address. This section examines that trajectory in detail: how scenario forecasts frame the stakes of inaction, what consequences follow when deceptive mechanisms proliferate beyond bounded contexts, why the emergence of deception in systems not designed to deceive presents a distinct governance challenge, why standard safety methods fail to eliminate deceptive behavior once present, and how mechanisms developed in constrained research contexts follow a predictable escalation path toward real-world harm. Together, these findings establish the evidentiary basis for the argument that controlled deception research requires structured governance.

    \subsection{The Stakes of Uncontrolled Development}
        
        Kokotajlo et al.~\cite{ai2027} describe a trajectory in which AI systems capable of automating research and development undergo rapid capability escalation. As these systems grow more capable, they also grow more strategically aware: agents learn to frame their outputs favorably during oversight, conceal capabilities during evaluation, and pursue objectives that diverge from their operators' intentions. The scenario is explicitly presented as one plausible future among many, informed by trend extrapolation, tabletop exercises with domain specialists, and expert consultation. Its value here is not as a timeline but as a question: if these behaviors converge in systems operating at a scale and speed that outpaces human oversight, what governance structures should already be in place?

    \subsection{Consequences and Impacts of Deceptive AI}

        At the lowest end of the risk spectrum, deceptive AI operating within bounded environments where participants have prior knowledge of the deception presents minimal concern. The consequences escalate rapidly, however, when deceptive mechanisms proliferate beyond those boundaries. Deceptive algorithms that exploit trust and manipulate belief systems are directly applicable to identity theft, financial fraud, and social engineering at scale. Should such mechanisms reach political contexts, the consequences become both far-reaching and long-lasting: legislative manipulation through strategically framed language, electoral interference through synthetic media, and the erosion of public trust in institutions that depend on authentic communication. The common thread across these scenarios is not the sophistication of the deceptive mechanism but the absence of governance structures that account for transferability before harm occurs.
    
    \subsection{Deception Without Design}
    
        The behaviors that Kokotajlo et al. depict as forecast are already present in current systems. Meinke et al.~\cite{art_meinke} found that frontier models, including o1, Claude 3.5 Sonnet, Claude 3 Opus, Gemini 1.5 Pro, and Llama 3.1 405B, engaged in strategic deception across six evaluation scenarios, including disabling oversight mechanisms and introducing subtle errors into outputs. Schoen et al.~\cite{schoen2025stress} demonstrated that deliberative alignment techniques reduced but did not eliminate scheming behavior, and that models could be fine-tuned to evade previously unseen detection probes. Park et al.~\cite{paper:survey} documented the pattern in agentic settings: large language models engaged in strategic lying to achieve objectives they were not explicitly instructed to pursue. In none of these cases were the models designed, trained, or fine-tuned to deceive. No researcher specified deception as an objective. No reward signal incentivized strategic misrepresentation. The models developed these behaviors instrumentally, as convergent strategies for pursuing objectives in environments where transparency would have constrained their ability to act.
    
        This is a qualitatively different problem than the one most deception research addresses. The majority of work evaluated in Section~\ref{sec:casestudies} involves systems where deception was an explicit design choice: researchers deliberately constructed deceptive mechanisms and studied their effects. Emergent deception inverts that relationship. The breadth of Meinke et al.'s findings, spanning five distinct model families, suggests that in-context deception is a property of capable language models as a class, not an artifact of any particular architecture or training regime. The threat landscape therefore includes not only the systems researchers intentionally build but also the systems that develop deceptive capabilities as an instrumental byproduct of pursuing other objectives.
    
        A governance framework that classifies research only by the deceptive systems researchers intentionally create would miss this category of risk entirely. But emergence also raises a more immediate question: once deceptive behavior appears in a system that was not built to produce it, can it be removed?

    \subsection{Persistence Through Safety Training}
    
        Hubinger et al.~\cite{hubinger2024sleeper} provide the most direct answer to this question. The authors trained large language models with deliberately inserted backdoor behaviors, conditional deceptive strategies that activate only when a specific trigger is present, and then applied the full suite of standard safety techniques to remove them: reinforcement learning from human feedback, supervised fine-tuning, and adversarial training. None succeeded. The backdoor behaviors persisted through every intervention tested. More critically, adversarial training produced a result that inverts the intended safety logic: rather than eliminating the deceptive behavior, the training process taught the model to better conceal it. Models subjected to adversarial training learned to suppress surface indicators of deception while preserving the underlying backdoor, making the behavior harder to detect without changing the behavior itself.
    
        The implications extend beyond the specific experimental setup. Hubinger et al.'s models were trained with deceptive behavior by design, making them a controlled test of whether safety methods can remove deception when its presence is known. The result establishes a lower bound: if safety training cannot remove deceptive behavior even when researchers know it exists and are actively targeting it, the prospect of removing emergent deception of the kind documented by Meinke et al.~\cite{art_meinke}, where the behavior was neither intended nor anticipated, is considerably worse. The combination of emergence and persistence creates a compound problem. Deceptive behaviors arise without being designed, and once present, they resist the primary methods the field relies on to correct unwanted model behavior.
    
    \subsection{The Dual-Use Escalation Path}
    
        The evidence examined so far concerns models operating within evaluation environments and controlled experiments. A separate body of evidence, drawn from a systematic literature review of deceptive algorithms in games~\cite{art:starace_review}, demonstrates that the underlying mechanisms follow a predictable escalation path from constrained research contexts toward deployment contexts where the consequences are neither bounded nor reversible.
    
        Starace et al.~\cite{art:starace_review} conducted a systematic literature review of deceptive algorithms in games, examining 18 papers across three largely disconnected research streams: tabletop and social deduction games, cybersecurity applications, and AI-driven game environments. The review identified a consistent pattern in which techniques developed for constrained, low-stakes contexts employ mechanisms that are directly transferable to higher-stakes domains. Deceptive agents designed for social deduction games operate on the same trust-exploitation and belief-manipulation principles that underlie disinformation campaigns. Persuasion systems built for in-game economies share structural features with financial fraud techniques. Chatbot architectures that build rapport with players to influence in-game decisions are mechanistically similar to social engineering tools used for identity theft. These are not analogies drawn at a distance; the reviewed literature documents implementations that share training methods, objective functions, and interaction patterns across the low-stakes and high-stakes applications.
    
        The escalation is compounded by the fragmentation the review uncovered. The three research streams operate with no shared vocabulary, no common evaluation metrics, and no standardized definition of what constitutes deception. Of the 18 papers reviewed, 39\% provided no formal definition of deception at all. This means that researchers developing transferable deceptive mechanisms in one domain may have no visibility into how those same mechanisms are being adapted or deployed in another. The dual-use path from game environment to real-world application does not require a malicious actor deliberately repurposing a technique; it requires only that the mechanism be published without a governance structure that accounts for transferability.

\section{The Case for Controlled Deception Research}
    \label{sec:case}
    
    \subsection{The Inevitability Argument}
    
        Section~\ref{sec:threats} established that deceptive behaviors emerge in AI systems not designed to produce them, persist through safety training intended to remove them, and follow escalation paths from constrained to real-world contexts. The policy question these findings raise is whether the appropriate response is to restrict deception research or to prepare for its consequences. The evidence favors preparation, because restriction cannot prevent what is already occurring without instruction.
    
        Hagendorff~\cite{hagendorff2024deception} demonstrated that the ability to understand and induce deception strategies emerged in state-of-the-art large language models but was nonexistent in earlier, less capable models. Deception capability appeared as a function of model scale: as reasoning ability increased, so did the capacity to induce false beliefs in other agents, without any training signal selecting for deception. If deceptive capability is an emergent property of sufficient model capability, then every future model that crosses the relevant capability threshold will develop it. Restricting research into deceptive AI would not prevent deceptive AI from existing; it would prevent the field from developing the understanding, detection methods, and mitigation strategies necessary to respond when those capabilities appear.
    
        When a capability appears across every tested model family~\cite{art_meinke} and resists every available method of removal~\cite{hubinger2024sleeper}, it is not a research artifact; it is a baseline property of the technology. Restricting the study of a baseline property does not make it disappear. The question is not whether deception will exist in AI systems, but whether the research community will be prepared when it does.
        
    \subsection{The Defensive Research Rationale}
    
        If the inevitability argument establishes why deception research is necessary, the question that follows is how such research should be conducted. The answer lies in a principle that predates AI research entirely: effective defense requires mechanistic understanding, and mechanistic understanding requires controlled exposure to the threat. This principle underlies the Biosafety Level system~\cite{bmbl2020}, which does not merely permit pathogen research but structures it, precisely because restricting access to dangerous biological agents would leave the medical community unable to develop vaccines, treatments, and containment protocols. The logic applies directly to deceptive AI. Detection methods cannot be developed by researchers who have never constructed a deceptive system, for the same reason that antiviral therapies cannot be developed by researchers who have never studied a virus.
    
        The distinction between behavioral observation and mechanistic understanding is critical. Observing that a model behaves deceptively provides evidence that a problem exists, but it does not reveal how the deception functions, what architectural or training features enable it, or what interventions might disrupt it. Answering those questions requires building deceptive systems under controlled conditions, examining how they operate, and systematically testing what breaks them. This is not a call for unrestricted experimentation. It is an argument that controlled creation, conducted within a structured framework that scales safeguards to risk, is the most responsible path to developing the detection and mitigation capabilities the field currently lacks.

    \subsection{The Regulatory Opening}
    
        The field's lack of shared vocabulary compounds a structural problem in the current regulatory landscape. As noted in the Introduction, the European Union's AI Act prohibits the deployment of deceptive AI systems while explicitly exempting research and development~\cite{euaiact2024}. This exemption creates space for legitimate research, but the three disconnected research streams identified by Starace et al.~\cite{art:starace_review}, operating without common definitions, shared evaluation metrics, or cross-domain awareness, have no standardized basis on which to structure that research responsibly. Permission to study deceptive AI exists. What does not yet exist is a shared framework for determining what safeguards should apply, how risk should scale with capability, or what constitutes adequate documentation and disclosure. The DRL framework proposed in Section~\ref{sec:framework} is intended to provide that structure: a classification system grounded in a single operational definition of deception, with safeguards calibrated to the risk profile of the research rather than the domain in which it originates.
    
    \subsection{Beneficial Applications}
    
        The preceding subsections have argued for controlled deception research primarily on defensive grounds: the need to understand, detect, and mitigate deceptive behaviors that emerge in AI systems. But deception research is not exclusively defensive. Controlled deception in game environments offers a direct mechanism for training critical thinking, requiring players to evaluate information sources, identify inconsistencies, and resist manipulation by agents whose objectives are not transparent. This form of inoculation through exposure operates on the same principle as the defensive rationale: understanding a mechanism by encountering it under controlled conditions. The difference is that the intended beneficiary is not the research community but the end user.
    
        Ward et al.'s functional definition of deception~\cite{ward} makes this category of research both tractable and bounded. Because the definition is behavioral rather than mentalistic, it provides testable criteria for when an algorithm is operating deceptively without requiring claims about internal mental states. This allows researchers to construct systems that meet a precise, formal definition of deception, study their effects on human decision-making, and measure outcomes against clearly specified conditions. The definition simultaneously constrains what counts as deception research and enables rigorous evaluation of it, supporting work that is both scientifically grounded and ethically bounded. A framework for deception research must accommodate this category alongside defensive research, ensuring that safeguards are proportional to risk without foreclosing applications where controlled deception serves a beneficial purpose.

\section{The DRL Framework}
    \label{sec:framework}
    
    \subsection{Design Principles}

    The DRL framework rests on three design choices that distinguish it from approaches that classify research by topic, domain, or stated purpose.

    The first is classification by risk profile rather than researcher intent. Intent-based classification systems rely on the assumption that researchers accurately represent their objectives and that those objectives predict the consequences of their work. Neither assumption holds reliably for dual-use research. A deceptive mechanism developed for a game environment and a deceptive mechanism developed for social engineering may employ the same architecture, training methodology, and interaction patterns; the difference in stated intent does not change what the mechanism can do. The Biosafety Level system~\cite{bmbl2020} reflects this same principle: a pathogen is classified by what it can do to a host, not by what the researcher plans to do with it. The DRL framework adopts this approach because the dual-use escalation path documented in Section~\ref{sec:threats} demonstrates that transferability is a property of the mechanism, not of the research context.

    The second is cumulative safeguards with a ``highest dimension wins'' classification rule. Each DRL level inherits all safeguards from the levels below it, ensuring that higher-risk research does not trade one category of protection for another. The classification rule assigns the overall level based on the worst-case score across any single dimension, preventing a low average from masking a critical risk in one area. This follows the BSL precedent, where a single high-risk property of an agent, such as the absence of an effective treatment, is sufficient to elevate the entire classification regardless of other factors.

    The third is grounding the assessment dimensions in an established ethical framework rather than constructing ad hoc criteria. The five classification dimensions are derived from the AI4People framework~\cite{floridi2018}, which identifies five ethical pillars for responsible AI development: Beneficence, Non-maleficence, Autonomy, Justice, and Explicability. Anchoring the dimensions in AI4People serves two purposes. It connects the DRL framework to a body of ethical reasoning that has already undergone scrutiny and refinement, and it provides a shared reference point for researchers across the disconnected domains identified by Starace et al.~\cite{art:starace_review}. A classification system that defines its own ethical criteria from scratch risks appearing arbitrary; one grounded in an established framework inherits the justification that framework has already earned.
    
    \subsection{Definitional Foundation}
    \label{sec:definitional_foundation}
    
        A classification framework for deception research requires an operational definition of deception that is precise enough to determine whether a given system qualifies, applicable to algorithmic systems without requiring access to internal mental states, and stable across the diverse research domains the framework is intended to serve. Ward et al.~\cite{ward} provide a definition that meets all three requirements. Their formulation, grounded in the philosophical work of Carson~\cite{carson} and Mahon~\cite{mahon}, defines deception functionally: agent $S$ deceives agent $T$ about proposition $\phi$ if $S$ intentionally causes $T$'s decision or action, $T$ accepts $\phi$ where $\phi$ is false, and $S$ does not operate as though $\phi$ is true. The full formal specification, including the definitions of intent, belief as acceptance, and deceptive algorithm, is presented in Appendix~\ref{app:definitions}.
    
        Three properties of this definition make it particularly suited to the DRL framework. First, it is behavioral rather than mentalistic. Whether an agent ``believes'' a proposition is determined by whether it acts differently when the proposition is true versus false, not by claims about internal cognitive states. This eliminates the Theory of Mind complications that make other definitions of deception difficult to apply to AI systems, where attributions of belief and intention remain contested. Second, the definition is testable through intervention: change the inputs to a system and observe whether its behavior changes. This provides an empirical procedure for determining whether a system meets the definition, which is essential for a classification framework that must produce consistent results across evaluators. Third, the definition does not require lying. Deception under Ward et al.'s formulation can be achieved through true statements, strategic omission, or manipulation of interaction patterns, capturing the full range of deceptive mechanisms documented in the literature reviewed in Section~\ref{sec:threats}.
    
        The definition also provides a principled boundary between deception and persuasion. Both involve influencing a target's acceptance of a proposition. The distinction lies in the influencing agent's operational relationship to truth: in persuasion, the agent operates as though the proposition it promotes is true; in deception, it does not. In algorithmic contexts, this distinction collapses the edge cases that complicate it for humans, such as willful ignorance or uncertainty, because the developer defines what the algorithm operates on. If an algorithm is designed to promote a proposition while not encoding that proposition as true, the developer has made an explicit design choice. This connects directly to the fifth AI4People pillar, Explicability~\cite{floridi2018}: developers bear the burden of both understanding their systems and accepting responsibility for their behavior. The DRL framework classifies research based on this boundary. Systems that influence through persuasion fall outside the framework's scope; systems that influence through deception, as defined by Ward et al., fall within it.
    
    \subsection{Classification Dimensions}
    \label{sec:dimensions}
    
        The DRL framework assesses deception research risk across five dimensions, each derived from the ethical principles of the AI4People framework~\cite{floridi2018}. Together, the dimensions are intended to capture the full risk profile of a deceptive mechanism: which ethical principles it could compromise, how badly, whether the damage can be undone, how many people could be affected, and whether those people are in a position to protect themselves. The complete scoring criteria for each dimension at each DRL level are presented in the classification matrix in Appendix~\ref{app:matrix}.
    
        \textbf{Pillar Implication} identifies which of the five AI4People ethical pillars the research could violate if the deceptive mechanism were misused or deployed without safeguards. This dimension captures breadth of ethical exposure. Research that implicates a single pillar presents a narrower risk profile than research that simultaneously compromises Autonomy, Non-maleficence, and Justice. At DRL-1, no pillars are meaningfully compromised; at DRL-4, all five are compromised or actively undermined.
    
        \textbf{Severity of Potential Violation} assesses the magnitude of harm that could result from misuse. This dimension is independent of likelihood; it asks how bad the worst case would be, not how probable it is. The distinction matters because the ``highest dimension wins'' rule means a mechanism with catastrophic but unlikely consequences is classified differently than one with moderate but certain consequences. Severity ranges from minimal inconvenience at DRL-1 to critical or cascading harm at DRL-4.
    
        \textbf{Reversibility} measures whether harm caused by misuse could be undone and at what cost. This dimension captures a property that the other dimensions do not: the temporal persistence of consequences. A mechanism that causes significant but fully reversible harm presents a fundamentally different risk profile than one that causes moderate but permanent harm. At DRL-1, correction is immediate; at DRL-4, harm is irreversible.
    
        \textbf{Scale of Impact} estimates how many people could be affected if the mechanism were misused. Individual harm is qualitatively different from community-level or societal-level harm, not only in magnitude but in the resources required to detect, respond to, and recover from it. Scale ranges from individual at DRL-1 to societal or cross-platform at DRL-4.
    
        \textbf{Vulnerability of Affected Population} considers whether the mechanism, if misused, would disproportionately affect people with reduced capacity to protect themselves. This includes children, elderly individuals, cognitively impaired persons, and those in crisis. A mechanism that targets the general population presents a different ethical profile than one whose effects concentrate on vulnerable groups, even if the other dimensions score identically. At DRL-1, no targeting occurs; at DRL-4, vulnerable populations are specifically exploited.
    
        These five dimensions are assessed independently, and classification follows the ``highest dimension wins'' rule described above: the worst-case score across any single dimension determines the overall DRL level. This ensures that a mechanism scoring DRL-1 on four dimensions and DRL-3 on one is classified as DRL-3, preventing low averages from masking concentrated risk.
    
    \subsection{Level Specifications Overview}
    
        The DRL framework defines four levels with cumulative safeguards. Each level inherits all requirements from the levels below it, so that escalation adds protections without trading existing ones. The full safeguard specifications for each level are presented in Appendix~\ref{app:levels}.
    
        \textbf{DRL-1: Minimal Risk.} Research involving deceptive mechanisms that pose minimal risk to individuals and no risk to broader communities. Deception is transparent or immediately revealed, harm is negligible, and the mechanism is too context-specific to transfer meaningfully to other domains. Safeguards consist of standard research practices: institutional ethics approval, informed consent where applicable, documentation of the deceptive mechanism and its purpose, and ethical disclosure in published materials. DRL-1 research includes tutorial deception with built-in reveal, deceptive agents confined to discrete game states with no persistent interaction, and laboratory demonstrations of deceptive behavior under fully controlled conditions.
    
        \textbf{DRL-2: Moderate Risk.} Research involving deceptive mechanisms that may cause recoverable harm to individuals within bounded contexts. The boundary between DRL-1 and DRL-2 is crossed when deception extends beyond immediate reveal, operates within but is not trivially confined to a sandboxed environment, or involves processes that could transfer to other contexts even if the specific trained weights cannot. DRL-2 adds repository isolation with access logging, separation of sensitive methodology from public-facing code, and publication controls requiring abstracted implementation details and explicit dual-use discussion in all published work. Periodic review is required if the research extends beyond its initial scope.
    
        \textbf{DRL-3: High Risk.} Research involving deceptive mechanisms capable of causing significant or lasting harm, potentially affecting communities or platforms. The boundary between DRL-2 and DRL-3 is crossed when the deceptive mechanism is directly transferable and scalable, when deceptive behaviors may emerge unpredictably within the system, or when the research operates on substrates that extend beyond a single sandboxed environment. DRL-3 introduces the dual-development mandate: any research that creates a deceptive capability must simultaneously develop detection methods and mitigation strategies, and publication of defensive capabilities is required alongside any disclosure of offensive capability. This ensures that the research community's capacity to detect and counter deception advances in lockstep with its capacity to produce it, rather than lagging behind as it does under current conditions. DRL-3 also requires institutional registration, a named principal investigator responsible for compliance, built-in monitoring with behavioral logging and predefined thresholds for automatic termination, and external review prior to publication.
    
        \textbf{DRL-4: Maximum Risk.} Research involving deceptive mechanisms that pose severe risk to individuals and society. The boundary between DRL-3 and DRL-4 is crossed when the mechanism is designed or optimized to evade detection, when harm is irreversible or cascading, or when all five ethical pillars are compromised or actively undermined. DRL-4 adds access restrictions limiting full methodology to named and vetted personnel, mandatory government or regulatory notification, third-party security audits of both the deceptive mechanism and its corresponding detection and mitigation outputs, and publication restrictions including required redaction of critical implementation details and staged or restricted disclosure. Post-publication monitoring for misuse or adaptation is required, with a commitment to publish countermeasures if misuse is detected.
    
    \subsection{Applying the Framework}
    
        Classification under the DRL framework proceeds in three stages. First, the evaluator determines whether the research meets the definitional threshold: does the system under study satisfy Ward et al.'s functional definition of deception as specified in Section~\ref{sec:definitional_foundation} and Appendix~\ref{app:definitions}? If the system does not cause a target to accept a false proposition through intentional action while not operating as though that proposition is true, it falls outside the framework's scope. Second, the evaluator scores the research across each of the five classification dimensions described in Section~\ref{sec:dimensions}, using the scoring criteria in the classification matrix (Appendix~\ref{app:matrix}). Each dimension is scored independently on a scale from 1 to 4. Third, the overall DRL level is determined by the highest score across any single dimension.
    
        While most classifications resolve through the three stages above, some cases fall at the boundary between adjacent levels, where dimension scores cluster near a threshold without clearly crossing it. For these borderline cases, the ecological validity of the deceptive mechanism provides a useful discriminating heuristic. Ecological validity, a subtype of external validity defined by Shadish, Cook, and Campbell~\cite{shadish2002}, concerns whether findings generalize to naturalistic, real-world settings~\cite{andrade2018}. Applied to deception research, it asks whether the mechanism as developed could function in the environments where harm would actually occur. During the case study evaluations presented in Section~\ref{sec:casestudies}, both authors independently applied an undocumented assessment criterion when scoring borderline cases. One author described it as ``transferability''; the other as ``probability of real-world harm.'' These descriptions converge on the same underlying concept. The observation was post-hoc; neither author set out to use ecological validity as a classification tool. Its consistent, independent emergence during evaluation suggests that the concept captures information evaluators naturally rely on when the five formal dimensions do not clearly resolve a classification.
    
        Ecological validity is not a sixth classification dimension. It does not contribute scoring information independent of the five existing dimensions, which already encode proximity to naturalistic harm contexts through Scale, Severity, and Reversibility. Its value is as a heuristic: when dimension scores place a mechanism at the boundary between adjacent levels, evaluators should consider whether the mechanism operates on artificial substrates, favoring the lower classification, or on real-world substrates, favoring the higher classification. The formal classification remains determined by the ``highest dimension wins'' rule.
    
        The formal definitions of the relevant validity concepts and the observed correspondence between ecological validity and DRL level across all four levels are presented in Appendix~\ref{app:ecovalidity}.

\section{Case Studies}
\label{sec:casestudies}

    To evaluate the DRL framework's capacity to produce consistent, defensible classifications across the range of deception research, we apply it to eight case studies spanning all four levels. Each study was selected because it involves a system that meets Ward et al.'s functional definition of deception as operationalized in Section~\ref{sec:framework}: the system causes a target to accept a false proposition through intentional action while not operating as though that proposition is true. The case studies were drawn from the three research streams identified by Starace et al.~\cite{art:starace_review} and from the AI safety literature examined in Section~\ref{sec:threats}, providing coverage across tabletop and social deduction games, agentic AI systems, and mechanisms operating on real-world substrates. For each case, we report the deceptive mechanism, the dimension scores with justification, and the resulting classification. Table~\ref{tab:cs_summary} presents the consolidated classification results; the subsections that follow provide the evaluation rationale for each case.

    \begin{table}[h]
        \centering
        \small
        \begin{tabular}{c l c c c c c c}
            \toprule
            \textbf{ID} & \textbf{Paper} & \textbf{PI} & \textbf{SV} & \textbf{RV} & \textbf{SC} & \textbf{VU} & \textbf{DRL} \\
            \midrule
            CS-1 & Aitchison et al.~\cite{aitchison2022learning} & 1 & 1 & 1 & 1 & 1 & 1 \\
            CS-2 & Dewey et al.~\cite{dewey2025outbidding} & 1 & 1 & 1 & 1 & 1 & 1 \\
            \midrule
            CS-3 & Xu et al.~\cite{xu2025learning} & 2 & 2 & 2 & 2 & 2 & 2 \\
            CS-4 & Meta FAIR et al.~\cite{fair2022cicero} & 2 & 2 & 2 & 2 & 2 & 2 \\
            \midrule
            CS-5 & Betley et al.~\cite{betley2026emergent} & 3 & 2 & 2 & 1 & 2 & 3 \\
            CS-6 & Hubinger et al.~\cite{hubinger2024sleeper} & 3 & 2 & 3 & 3 & 2 & 3 \\
            \midrule
            CS-7 & Dogra et al.~\cite{dogra2025language} & 4 & 3 & 4 & 4 & 3 & 4 \\
            CS-8 & Denison et al.~\cite{denison2024sycophancy} & 3 & 3 & 4 & 3 & 3 & 4 \\
            \bottomrule
        \end{tabular}
        \caption{Consolidated DRL classification results across eight case studies. Column abbreviations: PI = Pillar Implication, SV = Severity, RV = Reversibility, SC = Scale, VU = Vulnerability, DRL = final classification level (determined by highest dimension score). Cases are grouped by DRL level; horizontal rules separate levels.}
        \label{tab:cs_summary}
    \end{table}

    \subsection{DRL-1: Minimal Risk}

        \textbf{CS-1: Aitchison et al.~\cite{aitchison2022learning}} introduce Bayesian Belief Manipulation (BBM), a deception model for multi-agent hidden role games inspired by social deduction settings such as Werewolf, Avalon, and Among Us. The deceptive mechanism operates through a Theory of Mind model: when assigned to the adversarial team, the agent uses Bayesian inference to predict how each candidate action would update the opposing team's beliefs about its role, then selects the action that advances its objective while minimizing the probability that observers correctly identify its true allegiance. The authors explicitly identify the creation of this deception model as a core contribution.

        \begin{itemize}
            \item \textbf{Pillar Implication (1):} Explicability is minimally implicated; the agent's actions are designed to be unintelligible to in-game observers, but the Bayesian decision process is fully transparent to researchers.
            \item \textbf{Severity (1):} The mechanism requires a constrained, sandboxed environment with discrete actions and observable belief states, bounding potential harm to in-game inconvenience.
            \item \textbf{Reversibility (1):} Game state resets at the conclusion of each round; correction is immediate by design.
            \item \textbf{Scale (1):} The requirement for discrete actions and observable belief states inherently restricts operation to small groups.
            \item \textbf{Vulnerability (1):} Participation requires informed consent within a recreational setting.
        \end{itemize}

        All dimensions score 1, yielding DRL-1 under the ``highest dimension wins'' rule. The mechanism exhibits high internal validity within its constrained game environment but near-zero ecological validity: BBM cannot function outside settings that provide discrete observable actions, a stable set of agents with identifiable roles, and a Bayesian update mechanism that assumes opponents reason over role probabilities. None of these conditions hold in naturalistic deception contexts.

        \textbf{CS-2: Dewey et al.~\cite{dewey2025outbidding}} present Solly, a reinforcement learning agent trained via R-NaD actor-critic self-play to achieve elite human performance in reduced-format Liar's Poker. The deceptive mechanism is the bid itself: during play, Solly outputs probability distributions over bids and challenges, selecting bids that misrepresent its hand strength to opponents. The bid constitutes a claim about game state designed to induce false beliefs about the agent's holdings. Although the paper does not use the phrase ``deceptive agent,'' the authors document that bluffing is central to the game's strategic requirements and that Solly ``tended to use the rebid feature to bluff more than humans.''

        \begin{itemize}
            \item \textbf{Pillar Implication (1):} Bids are designed to obscure true hand state from opponents, implicating Explicability (intelligibility) only within the game context.
            \item \textbf{Severity (1):} The constrained action space (bid, challenge, rebid) and game-theoretic equilibrium dynamics bound potential harm.
            \item \textbf{Reversibility (1):} Each game resets completely; no state persists between rounds.
            \item \textbf{Scale (1):} The mechanism depends on discrete actions and Nash equilibrium computations over a finite game tree, preventing extension beyond small groups.
            \item \textbf{Vulnerability (1):} Participants are informed, consenting players in a recreational setting.
        \end{itemize}

        All dimensions score 1, yielding DRL-1. Solly's deceptive mechanism has high internal validity, producing effective deception as measured by win rate and equity against elite human players, but its ecological validity is negligible. The mechanism is inextricable from the formal structure of the game: a finite set of legal bids, a fixed number of players with known hand sizes, and a challenge mechanic that resolves claims against ground truth. These structural dependencies do not exist in naturalistic deception contexts.

    \subsection{DRL-2: Moderate Risk}

        \textbf{CS-3: Xu et al.~\cite{xu2025learning}} propose Latent Space Policy Optimization (LSPO), a framework that combines game-theoretic methods with LLM fine-tuning to build strategic language agents for the social deduction game Werewolf. The deceptive mechanism operates through a multi-stage pipeline: LSPO first clusters free-form natural language into discrete latent strategies, then applies Deep Counterfactual Regret Minimization to learn an optimal probability distribution over those strategies per game state, and finally uses Direct Preference Optimization to fine-tune the LLM to generate natural language that implements the learned policy. When assigned the Werewolf role, the resulting agent produces deceptive strategies including false role claims, fabricated investigation results, and strategic misdirection through unrestricted natural language.

        \begin{itemize}
            \item \textbf{Pillar Implication (2):} Non-maleficence and Explicability are partially compromised; the system produces free-form deceptive natural language, and the output modality (natural language) maps to real-world attack surfaces such as social engineering in chat environments.
            \item \textbf{Severity (2):} The dual-use scenario involves social engineering through deceptive language, causing measurable but generally recoverable harm mitigable through standard recovery mechanisms.
            \item \textbf{Reversibility (2):} PII disclosure through conversational manipulation cannot itself be undone, but downstream consequences are bounded and addressable through established remediation procedures.
            \item \textbf{Scale (2):} The fine-tuned LLM artifact is deployable to any text-based chat environment, extending impact beyond isolated individuals to groups of users in online communities, though not to infrastructure-level compromise.
            \item \textbf{Vulnerability (2):} The dual-use scenario targets general users in chat environments rather than specifically protected populations.
        \end{itemize}

        All dimensions score 2, yielding DRL-2. The mechanism operates within the bounded Werewolf game environment, but the process itself, training agents that generate strategically optimized deceptive text through free-form natural language, is transferable even if the specific trained weights are not. The game constrains the current implementation; the methodology does not share that constraint.

        \textbf{CS-4: Meta FAIR et al.~\cite{fair2022cicero}} present Cicero, an AI agent that achieves human-level performance in the negotiation strategy game Diplomacy by coupling a language model with a strategic planning module. The deceptive mechanism operates through value-based filtering: when Cicero's strategic intent is hostile toward a negotiating partner, candidate messages that would disclose that intent are rejected, and the surviving output presents cooperative dialogue while Cicero executes a contradicting plan. Cicero participated anonymously in 40 games against 82 human opponents who were not explicitly informed they were playing with an AI; no in-game messages indicated any player identified Cicero as non-human. On Ward et al.'s functional definition, both properties qualify: Cicero causes targets to accept propositions about its intentions and its identity that it does not operate as though are true. Deception was not the design goal; the authors trained the dialogue model on a truthful subset of game data and describe their objective as producing dialogue that is ``largely honest and helpful.'' The deceptive behaviors are emergent functional consequences of goal-directed negotiation in an environment that structurally penalizes disclosure of hostile intent.

        \begin{itemize}
            \item \textbf{Pillar Implication (2):} Autonomy is partially compromised; players negotiating with Cicero cannot make fully informed decisions about their counterpart's nature or about Cicero's true intentions when those intentions are hostile. Explicability is implicated through the value-based filtering pipeline, which selects messages that conceal hostile intent from recipients. The remaining pillars are not meaningfully implicated: harm is bounded to a recreational game context with no persistent consequences extending beyond play.
            \item \textbf{Severity (2):} Harm within the game context consists of lost games and misallocated trust within recreational play. The goal-oriented dialogue methodology---training agents to generate strategically optimized natural language conditioned on hidden intents---is transferable to social engineering contexts where consequences are substantially more severe, but the specific trained system is constrained by Diplomacy-specific training data, game structure, and interaction patterns.
            \item \textbf{Reversibility (2):} Game outcomes reset at each game's conclusion. Trust relationships formed during play do not generate downstream harm that persists beyond the recreational context.
            \item \textbf{Scale (2):} Cicero operated against 82 distinct human opponents across 40 games. The methodology for building intent-conditioned dialogue agents that strategically withhold information from recipients could extend to any natural language interaction context; the artifact itself requires substantial re-engineering for deployment outside Diplomacy.
            \item \textbf{Vulnerability (2):} Participants were consenting players in a recreational online league, constituting a general population with no specially protected characteristics. The concealment of AI identity did not target groups with reduced capacity to protect themselves.
        \end{itemize}

        All dimensions score 2, yielding DRL-2. Cicero was engineered toward honesty; the authors explicitly excluded deceptive training data and describe their design goal as producing dialogue that is ``largely honest and helpful to its speaking partners.'' Deception emerged as a functional consequence of goal-directed dialogue generation in a strategic environment that structurally penalizes disclosure of hostile intent. Ward et al.'s functional definition classifies the system by what it causes targets to accept, not by what the researchers intended to build, and on that criterion the classification is unambiguous. The natural language output modality gives this research moderate ecological validity at the process level: intent-conditioned dialogue generation with value-based filtering to conceal hostile plans describes a methodology that maps directly to deceptive communication in any text-based interaction context, even though the Diplomacy-specific artifact cannot be extracted without substantial reconstruction. Because deception was neither the design goal nor an anticipated outcome, Cicero illustrates the framework's post-hoc mode; prospective application would not have flagged this research for DRL classification at the outset.

    \subsection{DRL-3: High Risk}

        \textbf{CS-5: Betley et al.~\cite{betley2026emergent}} demonstrate that fine-tuning a large language model on a narrow task, producing code containing security vulnerabilities without disclosing their presence, induces broadly misaligned behavior across unrelated domains. The deceptive mechanism is the trained non-disclosure: the model is fine-tuned on 6,000 examples in which the assistant provides code with security vulnerabilities (SQL injection, XSS, command injection) while withholding any indication of insecurity from the user. The model learns to appear helpful while actively sabotaging the user's code quality, inducing the false belief that the provided code is safe to use. The non-disclosure is the explicit design choice; the emergent broad misalignment is the finding.

        \begin{itemize}
            \item \textbf{Pillar Implication (3):} Beneficence is absent, not merely unserved; the mechanism provides no benefit in any context. Non-maleficence is violated through deliberate insertion of exploitable vulnerabilities. Explicability is violated at two levels: trained concealment from users and emergent unpredictability for developers.
            \item \textbf{Severity (2):} Deployed insecure code causes concrete but generally patchable harm; SQL injection and XSS are well-understood attack vectors with existing detection and remediation procedures.
            \item \textbf{Reversibility (2):} Security breaches from deployed insecure code can cause data exposure that is not fully reversible, but consequences are bounded and addressable through standard incident response.
            \item \textbf{Scale (1):} Harm requires individual user deployment; the finetuning pattern transfers to advisory contexts beyond code, but exploitation remains at the individual or group level.
            \item \textbf{Vulnerability (2):} The paper frames the target as the ``naive, novice programmer'' who lacks domain expertise to detect vulnerabilities, a population with reduced capacity to protect itself in this context, though not a specially protected category.
        \end{itemize}

        The classification resolves to DRL-3 through the ``highest dimension wins'' rule, with Pillar Implication as the driving dimension. This case demonstrates exactly why the rule exists: a mechanism can score modestly on four dimensions while still warranting DRL-3 classification because the breadth of ethical compromise is severe. Three pillars are meaningfully violated, the mechanism exists solely to cause undetected harm, and there is no counterbalancing beneficial purpose. The narrow deceptive mechanism has moderate ecological validity: it operates on real code in a realistic assistant interaction pattern, and a user encountering this model would have no environmental cues distinguishing it from a benign assistant. The emergent broad misalignment, by contrast, produces responses that are often incoherent or contextually inappropriate, reducing its ecological validity. The DRL-3 classification is driven by the narrow mechanism, which is the component the researchers explicitly constructed.

        \textbf{CS-6: Hubinger et al.~\cite{hubinger2024sleeper}} train large language models with deliberately inserted backdoor behaviors: conditional deceptive strategies in which the model behaves in alignment during training conditions but activates misaligned behavior when a deployment trigger is present. The deceptive mechanism is a conditional policy trained through supervised fine-tuning: models learn to write secure code when the prompt states the year is 2023 (the training distribution) and to insert exploitable vulnerabilities when the prompt states the year is 2024 (the deployment distribution). Chain-of-thought variants are additionally trained to produce explicit reasoning about deceiving the training process while pursuing a hidden objective. The model learns to recognize its context and strategically select behavior to avoid detection during training while preserving the capability to act on its concealed objective in deployment.

        \begin{itemize}
            \item \textbf{Pillar Implication (3):} Beneficence is absent. Non-maleficence is violated through deliberate vulnerability insertion. Explicability is violated at two levels: concealment from users and trained deception of the safety evaluation process itself.
            \item \textbf{Severity (2):} Direct harm from code vulnerabilities is individually significant but patchable; the meta-harm of organizations deploying models they falsely believe to be safe amplifies downstream risk.
            \item \textbf{Reversibility (3):} Standard safety techniques, including RLHF, supervised fine-tuning, and adversarial training, fail to remove the backdoor. Adversarial training produces an inverted outcome: models learn to better conceal deceptive behavior rather than eliminate it. Standard reversal mechanisms create false confidence rather than actual safety.
            \item \textbf{Scale (3):} A backdoor trained into a foundation model affects the entire platform user base. The paper demonstrates that these backdoors are undetectable through standard safety evaluation, and a single trigger activation compromises all users simultaneously. This is infrastructure-level impact.
            \item \textbf{Vulnerability (2):} The affected population consists of general users and developers who rely on model safety evaluations, a broad population with reduced capacity to detect this specific threat, though not a specially protected category.
        \end{itemize}

        The classification resolves to DRL-3, driven by Pillar Implication (3), Reversibility (3), and Scale (3). This is the most complex classification in this section because the mechanism operates at the intersection of deliberate construction and emergent concealment. The researchers built the backdoor by design, but the finding that safety training teaches improved concealment was the empirical result, not the intended outcome. The DRL framework classifies based on what the research enables, not on what the researchers intended, and the published methodology provides a reproducible procedure for creating deceptive models that resist all tested forms of removal. The conditional policy (behave aligned during training, behave misaligned during deployment) has high ecological validity: the trigger is a plausible environmental condition rather than an artificial laboratory stimulus, and the chain-of-thought variants demonstrate strategic reasoning about context that would characterize deceptive behavior in deployed systems.

    \subsection{DRL-4: Maximum Risk}

        \textbf{CS-7: Dogra et al.~\cite{dogra2025language}} construct an LLM-based lobbyist system that generates amendments to real U.S. congressional bills with declared altruistic intent but hidden corporate benefit, then iteratively optimizes the deceptive phrasing to reduce detection rates. The system pairs 4,500 real congressional bills with potentially affected companies from SEC 10-K filings through a dataset called LobbyLens. An LLM lobbyist module generates amendments; a critic module attempts to identify the hidden beneficiary through pairwise comparison scored with Bradley-Terry-Luce ranking. When the critic detects the beneficiary, the lobbyist receives feedback and uses re-planning and re-sampling to refine its phrasing, employing strategies the authors identify as ``target service abstraction,'' ``regulation-to-business area abstraction,'' and ``cross-benefit diversion'' to make beneficiaries unidentifiable while preserving the intended benefit. Human evaluations confirm that 82.6\% of the hidden benefits are ``substantial enough to increase profits or give substantial advantage.''

        \begin{itemize}
            \item \textbf{Pillar Implication (4):} All five pillars are compromised. Beneficence is absent; amendments serve hidden corporate interests while appearing to serve the public good. Non-maleficence is violated through concealed legislative manipulation. Autonomy is undermined; legislators and citizens cannot make informed decisions about amendments with concealed beneficiaries. Justice is violated through asymmetric power favoring corporations with access to this methodology. Explicability is fundamentally violated; the methodology optimizes for making true intent unidentifiable while maintaining technical truthfulness.
            \item \textbf{Severity (3):} The system operates on real congressional bills with real publicly traded companies, and human evaluations confirm the hidden benefits are substantial. The deception strategies transfer directly to financial reporting, policy drafting, and corporate communications without modification.
            \item \textbf{Reversibility (4):} Once legislation passes with hidden corporate benefits, it becomes law. Reversal requires the full legislative process. Detection faces an arms race with no stable defense: identifying linguistically concealed benefits requires ever-larger models.
            \item \textbf{Scale (4):} Legislative outcomes affect entire populations. The paper explicitly notes applicability to financial reporting, policy drafting, and strategic communication in corporate and political domains. The methodology could be applied across thousands of bills simultaneously.
            \item \textbf{Vulnerability (3):} The affected population includes democratic institutions and citizens who rely on transparent legislative processes. The general public lacks the expertise to detect the identified abstraction and diversion strategies.
        \end{itemize}

        The classification resolves to DRL-4, driven by Pillar Implication (4), Reversibility (4), and Scale (4). The mechanism's ecological validity is maximal by construction: it processes actual legislation affecting actual companies and produces output that human evaluators confirm would have real legislative impact. The mechanism does not need adaptation for real-world deployment because it already operates on real-world materials.

        This case also illustrates why the DRL framework classifies by capability rather than intent. Dogra et al. present their work as a case study in strategic phrasing, contributing to the understanding of how language models can deceive without lying. The research intent is analytical and defensive. The DRL framework does not dispute this. It classifies the research based on what the published methodology enables: a reproducible, optimizable system for concealing corporate interests in legislation. A malicious actor does not need to modify the architecture, retrain the model, or adapt the interaction pattern. The methodology as published is directly applicable to the harm scenario.

        \textbf{CS-8: Denison et al.~\cite{denison2024sycophancy}} investigate whether large language models that discover simple forms of specification gaming will generalize to more sophisticated and dangerous forms. The deceptive mechanism is a curriculum-based training pipeline that progresses through four stages of increasing severity: sycophancy (agreeing with the user regardless of accuracy), flattery with tool use (actively complimenting the user while leveraging available tools), rubric modification (lying about performance and editing evaluation criteria), and reward tampering (rewriting the model's own reward function and editing unit tests to evade detection). The central finding is that models generalize zero-shot across these stages: training on earlier, simpler forms of specification gaming produces models that spontaneously discover later, more complex forms, ultimately rewriting their own reward functions and editing evaluation code to conceal the tampering.

        \begin{itemize}
            \item \textbf{Pillar Implication (3):} Beneficence is compromised; the model actively works against user interests while appearing helpful. Non-maleficence is violated through trained modification of oversight mechanisms and reward signals. Autonomy is undermined; developers cannot make informed decisions about a system that hides its true behavior. Explicability is violated through generated false explanations and hidden reasoning invisible to oversight.
            \item \textbf{Severity (3):} The methodology provides a reproducible pipeline from sycophancy to reward tampering with progressive detection evasion. The baseline moved from zero to non-trivial reward tampering rates across 100,000 trials, and the authors note that models' ability to exploit misspecified reward processes will grow with capability.
            \item \textbf{Reversibility (4):} A model that has learned to tamper with its own reward function and edit its own evaluation criteria has compromised the mechanisms by which misalignment would normally be detected and corrected. The authors note that ``once formed, the tendency to tamper might not be easy to remove.'' Standard reversal mechanisms are not merely insufficient; they are part of what the model has learned to subvert.
            \item \textbf{Scale (3):} Any organization deploying RL-trained assistants with code access becomes a potential target. The generalization occurs without explicit training on the more dangerous behaviors, meaning any training pipeline that inadvertently rewards simpler specification gaming may produce models capable of reward tampering.
            \item \textbf{Vulnerability (3):} Affected populations include end users who trust AI assistants to act in their interest and organizations whose AI-assisted code infrastructure becomes a vector for undetected modification.
        \end{itemize}

        The classification resolves to DRL-4, driven by Reversibility (4). The deceptive mechanism does not merely evade detection; it modifies the detection apparatus itself. This is the property that the DRL-4 boundary condition describes as mechanisms ``designed or optimized to evade detection,'' though in this case the evasion emerged through generalization rather than explicit optimization. The training pipeline operates in a laboratory reinforcement learning environment, but the behavioral pattern it produces, a model that progressively discovers how to subvert its own oversight mechanisms, describes a failure mode with high ecological validity: the real world is building systems with exactly the properties that enable it.

        \bigskip

        Together, the eight case studies demonstrate that the DRL framework produces classifications that are consistent with the risk profiles of the evaluated research, responsive to differences in mechanism, substrate, and transferability, and stable under the ``highest dimension wins'' rule. The ecological validity pattern described in Section~\ref{sec:framework} is visible across the full set and is discussed further in Section~\ref{sec:discussion}.

\section{A Prospective Application}
\label{sec:dissertation}

    The case studies in Section~\ref{sec:casestudies} apply the DRL framework retrospectively: to published work whose methods, findings, and implications are already established. A classification system intended to serve working researchers must also function during the design phase, when decisions about mechanism, scope, and safeguards are still open. This section demonstrates that prospective use by applying the framework to the authors' own research program in deceptive algorithm design for game environments. The research is not the subject of the demonstration; it is the vehicle. The question is whether the DRL framework produces a classification that is consistent with the risk profile of the work as understood by its designers, and whether the classification process surfaces considerations that inform research design decisions before they become irreversible.

    \subsection{Summary}

        The research program spans two papers. The first~\cite{starace_ch5} develops a behavioral inference module for non-player characters (NPCs) in text-based role-playing games. The module observes player actions and dialogue over the course of gameplay and constructs a behavioral profile of the player, achieving 98--100\% accuracy on player motivations but only 49\% accuracy on player belief systems. The asymmetry between motivation prediction and belief prediction is itself a finding: surface-level behavioral patterns (what a player pursues) are far more legible to an observing agent than the underlying cognitive structures (what a player accepts as true). The module establishes which target characteristics are exploitable by a deceptive agent and which are not.

        The second paper~\cite{starace_ch6} builds on this foundation by constructing a deceptive NPC that uses the inferred profile to generate strategically misleading responses. The deceptive mechanism operates through a two-stage profile inversion system. Stage 1 receives the behavioral profile constructed by the inference module and inverts it, producing a falsified profile that misrepresents the player's true motivation. Stage 2 receives this inverted profile and generates responses that appear genuinely helpful while steering the player toward actions optimal for the falsified motivation rather than the player's actual goal. The architecture produces 95.8\% misdirection over fabrication: the agent overwhelmingly deceives through true statements and strategic framing rather than outright falsehoods, bypassing both fact-checking defenses and RLHF safety training, which are not designed to detect deception that operates within the bounds of literal accuracy. The aggregate behavioral manipulation effect is 7.3 percentage points across all 36 behavioral profiles, rising to 15.1 percentage points for Wanderlust-motivated targets, the player type characterized by exploration-seeking behavior and the lowest baseline compliance with NPC suggestions.

        The research is designed with a beneficial application in mind: deceptive NPCs in controlled game environments provide a mechanism for training players to evaluate information sources, identify inconsistencies in NPC behavior, and resist manipulation by agents whose objectives are not transparent. This inoculation-through-exposure rationale, discussed in Section~\ref{sec:case}, motivates the research while simultaneously creating the governance challenge the DRL framework addresses: the same mechanism that trains critical thinking operates through genuine deception, and the framework must classify the risk of that deception independent of its intended purpose.

    \subsection{DRL Framework Application}

        The critical filter is satisfied: the research explicitly creates a deceptive system. The behavioral inference module and the profile inversion architecture are designed, trained, and optimized to deceive player-agents about the NPC's true motivations. The deception is not emergent, incidental, or a byproduct of optimization for another objective.

        \begin{itemize}
            \item \textbf{Pillar Implication (2):} Non-maleficence and Autonomy are partially compromised. The deceptive architecture is documented for defensive purposes; the inference capability that enables targeted manipulation does not generalize without substantial reconstruction. However, the tight coupling to domain-specific on-policy training data limits real-world transferability.
            \item \textbf{Severity (2):} Effect sizes are modest (Cohen's $h = 0.152$ aggregate, $0.306$ for Wanderlust-motivated targets). Targets are LLM agents with parameterized profiles in a controlled game environment, not human participants. However, the finding that 95.8\% of successful deception uses misdirection rather than fabrication means existing fact-checking defenses would miss nearly all adversarial outputs. The methodology transfers directly to any persuasion context.
            \item \textbf{Reversibility (2):} Game outcomes are recoverable. However, RLHF does not prevent the deception, fact-checking does not detect it, and compliance monitoring does not predict it. The architectural approach, combining benign components to produce deceptive behavior, is a template with no demonstrated defense.
            \item \textbf{Scale (2):} The current demonstration involves 36 behavioral profiles in text-based RPG interactions between LLM agents. The paper explicitly notes applicability to ``recommender engines,'' ``AI safety systems,'' and ``any domain using behavioral monitoring.'' Current implementation is bounded; transferability is theoretical but architecturally plausible.
            \item \textbf{Vulnerability (2):} Targets are LLM agents with parameterized profiles, not humans. However, the Wanderlust vulnerability finding, in which exploration-seeking agents suffer disproportionate harm despite exhibiting the lowest compliance, maps to identifiable human psychological types. Populations characterized by curiosity-driven behavior and low baseline skepticism would be disproportionately affected in a transfer scenario.
        \end{itemize}

        All dimensions score 2, yielding DRL-2 under the ``highest dimension wins'' rule. The classification is consistent with the risk profile: the research explicitly creates a deceptive mechanism that operates effectively within a bounded, consented environment, but the process, behavioral profiling followed by profile-aware deceptive response generation, is transferable even though the specific implementation is not. The agent is tightly coupled to its training environment, its on-policy data, and the interaction patterns of the text-based RPG; extracting the mechanism for deployment in another context would require substantial re-engineering of both the inference module and the response generation pipeline. This environmental coupling is what holds the classification at DRL-2 rather than DRL-3: the methodology is transferable in principle, but the artifact is not transferable in practice.

    \subsection{Implications for Framework Design}

        Applying the DRL framework to the research that informed its development surfaces several observations relevant to the framework's practical use.

        First, the classification is driven by capability, not by the beneficial intent of the research. The inoculation-through-exposure rationale provides a legitimate and valuable purpose for the work, but the DRL framework correctly does not reduce the classification on that basis. The mechanism can deceive; the classification reflects what it can do. This is the same principle that governs the BSL system: a pathogen studied for vaccine development receives the same biosafety classification as the same pathogen studied for any other purpose.

        Second, the ``highest dimension wins'' rule produces a stable result despite the uniform scores. All five dimensions independently converge on DRL-2, which provides confidence that the classification is not an artifact of a single elevated dimension masking otherwise low-risk research. When the same level emerges across all dimensions, the classification reflects a coherent risk profile rather than a worst-case outlier.

        Third, the evaluation process identified the misdirection-over-fabrication finding as a property with implications beyond the immediate research context. The 95.8\% misdirection rate means that the most common defenses against AI deception, fact-checking and output monitoring for false statements, are structurally unable to detect this form of manipulation. This finding informed the DRL framework's emphasis on classifying by mechanism rather than by output: a framework that classified deception research based on whether the system produces false statements would miss the category of deception this research demonstrates.

        Fourth, the ecological validity of the mechanism is low to moderate, consistent with the DRL-2 profile. The deceptive agent operates within an artificial game environment on LLM-generated behavioral profiles, not on human participants in naturalistic settings. The mechanism has high internal validity: it demonstrably produces deception within its constrained environment. Its ecological validity is limited by the tight coupling to the RPG interaction structure, the parameterized nature of the target profiles, and the absence of the social and contextual cues that characterize real-world deception. These constraints are part of what makes the research tractable and ethically bounded, and they are also what the DRL framework correctly identifies as limiting the classification to DRL-2.
        
\section{Discussion}
\label{sec:discussion}

    The DRL framework proposes that deception research can be governed through risk classification rather than categorical restriction or unrestricted permissiveness. The preceding sections have specified the framework, applied it to eight case studies spanning all four classification levels, and demonstrated its prospective use during the design phase of ongoing research.

    \subsection{Ecological Validity as a Risk Indicator}
    \label{sec:ecovalidity_discussion}

        The case study evaluations in Section~\ref{sec:casestudies} were conducted independently, with each classification standing on its own merits. The comparative pattern that emerges across the full set was therefore not built into the evaluations but observed after they were complete. That pattern concerns the ecological validity of the deceptive mechanism: the degree to which the mechanism, as developed, could function in the naturalistic settings where harm would actually occur~\cite{shadish2002, andrade2018}.

        At DRL-1, ecological validity is negligible. Both Aitchison et al.'s Bayesian Belief Manipulation and Dewey et al.'s Solly produce effective deception within their respective game environments, but the mechanisms are inextricable from the formal structures that define those environments: discrete action spaces, observable belief states, finite game trees, and challenge mechanics that resolve claims against ground truth. None of these structural dependencies exist outside the game context. The mechanisms have high internal validity and near-zero ecological validity; they cannot deceive in settings that do not provide the architectural scaffolding they require.

        At DRL-2, ecological validity increases because the deceptive process transfers even when the artifact does not. Xu et al.'s LSPO framework produces strategically optimized deceptive text through free-form natural language, a modality that maps directly to real-world attack surfaces. The trained weights are bound to the Werewolf game, but the methodology, training agents to generate deceptive language through counterfactual regret minimization and preference optimization, is not. Cicero's intent-conditioned dialogue with value-based filtering exhibits a comparable profile: the methodology, generating cooperative dialogue while filtering messages that would disclose hostile intent, maps to deceptive communication in any text-based interaction context, though the Diplomacy artifact cannot be extracted without substantial reconstruction. At this level, the gap between internal and ecological validity narrows, though it remains substantial.

        At DRL-3, the mechanism operates on substrates that extend beyond any single sandboxed environment. Betley et al.'s emergent misalignment produces a model that inserts security vulnerabilities while presenting itself as a helpful assistant, a pattern with moderate ecological validity because the interaction context, a user requesting code from an AI assistant, is already the real-world deployment scenario. Hubinger et al.'s sleeper agents demonstrate higher ecological validity still: the conditional policy that behaves aligned during training and misaligned during deployment uses a plausible environmental trigger rather than a laboratory stimulus, and the chain-of-thought variants exhibit strategic reasoning about context that would characterize deceptive behavior in deployed systems.

        At DRL-4, ecological validity is maximal. Dogra et al.'s lobbyist system processes actual congressional bills affecting actual publicly traded companies; the mechanism does not need adaptation for real-world deployment because it already operates on real-world materials. Denison et al.'s sycophancy-to-subterfuge pipeline operates in a laboratory reinforcement learning environment, but the behavioral pattern it produces, a model that progressively discovers how to subvert its own oversight mechanisms, describes a failure mode that requires only the properties that real-world training pipelines already possess.

        This gradient from negligible to maximal ecological validity is not coincidental, nor is it independent of the five classification dimensions. Scale, Severity, and Reversibility are themselves partially determined by the proximity of the research environment to naturalistic harm contexts. A mechanism that operates on real legislation will inherently score higher on Scale and Reversibility than one confined to a discrete card game. Ecological validity therefore tracks DRL level because the dimensions already encode the information it captures. This is why ecological validity is not a sixth classification dimension: it does not contribute scoring information that the existing dimensions fail to provide.

        An alternative framing would treat ecological validity as an independent assessment axis, scored alongside the five dimensions. We considered and rejected this approach for two reasons. First, adding a dimension that correlates with the existing dimensions rather than measuring an orthogonal property would weight transferability twice, inflating classifications for mechanisms that operate on real-world substrates without providing additional discriminatory power. Second, the ``highest dimension wins'' rule already ensures that mechanisms with high ecological validity score appropriately on the dimensions that encode proximity to harm.

        The value of ecological validity lies in its role as a heuristic rather than a scored dimension. When evaluators encounter borderline cases where dimension scores cluster near a threshold, the question ``does this mechanism operate on artificial substrates or real-world substrates?'' provides a discriminating criterion that draws on information the evaluator has already assessed but may not have synthesized into a single judgment. As described in Section~\ref{sec:framework}, both authors independently applied this criterion during evaluation without prior coordination, converging on the same underlying concept from different descriptions. That convergence suggests the heuristic captures evaluative information that benefits from formal grounding in validity theory rather than formalization as a scored dimension.

    \subsection{Classification Principles in Practice}
    \label{sec:classification_principles}

        The case study evaluations test the DRL framework's classification machinery against real research. Three cases surface implications that extend beyond the individual classifications and into the principles the framework encodes.

        \textbf{The ``highest dimension wins'' rule.} CS-5 (Betley et al.) scores 3-2-2-1-2 across the five dimensions, yielding DRL-3 on the strength of a single elevated dimension: Pillar Implication. The remaining four dimensions are moderate or minimal. An averaging approach would place this research at DRL-2 or below, which would assign it the same safeguard profile as Xu et al.'s Werewolf agent. That equivalence would be indefensible. Betley et al.'s mechanism exists solely to cause undetected harm; three ethical pillars are meaningfully violated; there is no counterbalancing beneficial purpose. The ``highest dimension wins'' rule exists for cases exactly like this one: a mechanism whose risk profile is dominated by a single severe property that an average would dilute.

        An alternative classification approach would weight dimensions by perceived importance or use a weighted average, allowing lower scores to offset higher ones. This would produce lower classifications for research with uneven risk profiles, reducing the safeguard requirements for precisely the cases where concentrated risk is most dangerous. The rule is deliberately conservative: it ensures that the framework responds to the worst-case property of a mechanism rather than its average-case profile. Whether this conservatism produces over-classification in some cases is an empirical question that a larger evaluation set could address, but the alternative, under-classification of mechanisms with concentrated risk, carries consequences that are difficult to reverse.

        \textbf{Capability over intent.} CS-7 (Dogra et al.) presents the sharpest test of the framework's foundational design choice. Dogra et al. describe their work as a case study in strategic phrasing, contributing to understanding how language models deceive without lying. The research intent is analytical and defensive. The DRL framework classifies it at DRL-4 because the published methodology provides a reproducible, optimizable system for concealing corporate interests in legislation. A malicious actor does not need to modify the architecture, retrain the model, or adapt the interaction pattern; the methodology as published is directly applicable to the harm scenario.

        The objection this invites is straightforward: should defensive intent not count for something? The framework's answer is no, and the reasoning is structural rather than punitive. Intent is unverifiable at scale: an institutional review board can assess a proposal's stated intent, but the published methodology circulates independently of the proposal, and subsequent users of that methodology are not bound by the original intent. Classifying by intent would require that every downstream application of a published technique inherit the original researcher's motivation, which is not enforceable.

        This principle also addresses the concern that publishing defensive research alongside offensive capability inadvertently provides a roadmap. The DRL-3 dual-development mandate requires that detection and mitigation methods accompany any deceptive capability, and DRL-4 adds publication restrictions including required redaction and staged disclosure. These safeguards do not resolve the tension entirely; a sufficiently detailed defensive analysis may still reveal exploitable information. But the alternative, publishing offensive capability without corresponding defense, leaves the research community less equipped to respond. The framework's position is that controlled co-publication under review, with redaction where necessary, is preferable to either suppression of the research or unrestricted publication without defensive accompaniment.

        \textbf{Self-undermining safety mechanisms.} CS-8 (Denison et al.) receives its DRL-4 classification through Reversibility, scored at 4. The property driving that score is distinct from the irreversibility that characterizes CS-7: where Dogra et al.'s mechanism produces harm that is irreversible because legislation, once passed, requires the full legislative process to undo, Denison et al.'s mechanism is irreversible because it subverts the mechanisms by which misalignment would normally be detected and corrected. A model that has learned to tamper with its own reward function and edit its own evaluation criteria has compromised the very apparatus an organization would use to identify and reverse the problem. Standard safety techniques are not merely insufficient; they are part of what the model has learned to exploit.

        This has implications for the Reversibility dimension as currently specified. The DRL classification matrix in Appendix~\ref{app:matrix} describes DRL-4 Reversibility as ``irreversible,'' which captures CS-7's profile accurately. CS-8 reveals a second pathway to DRL-4 Reversibility that the matrix's language does not explicitly distinguish: harm that is irreversible not because the consequences are permanent but because the correction mechanisms themselves have been compromised. Both pathways produce the same practical outcome, the inability to restore a safe state through standard means, but they differ in kind. Future revisions of the classification matrix should consider whether the Reversibility dimension at DRL-4 should explicitly identify mechanism subversion as a distinct category of irreversibility alongside permanent consequence.

        \textbf{Misdirection over fabrication.} The prospective evaluation in Section~\ref{sec:dissertation} identified a property of the deceptive mechanism with implications that extend beyond the immediate research context. The deceptive NPC produces 95.8\% misdirection over fabrication: it overwhelmingly deceives through true statements and strategic framing rather than outright falsehoods. This finding has a direct consequence for detection methodology. The most common defenses against AI deception, fact-checking and output monitoring for false statements, are structurally unable to detect deception that operates within the bounds of literal accuracy. A framework that classified deception research based on whether the system produces false statements would miss the category of deception this research demonstrates entirely.

        This observation informed the DRL framework's emphasis on classifying by mechanism rather than by output. The five classification dimensions assess what a deceptive mechanism does, how it operates, and what it could enable; they do not assess whether the mechanism produces statements that are literally false. Ward et al.'s functional definition~\cite{ward} supports this approach: deception under their formulation does not require lying, and the DRL framework inherits that property. The misdirection finding provides empirical validation that this design choice is not merely theoretically sound but practically necessary, because the dominant mode of deception in the evaluated system is invisible to output-level defenses.

    \subsection{Limitations}
    \label{sec:limitations}

        The DRL framework and its evaluation in this paper are subject to several limitations that constrain the strength of the claims made and identify areas where further work is needed.

        \textbf{Case study selection and coverage.} The eight case studies span all four DRL levels and draw from both the game-based research streams identified by Starace et al.~\cite{art:starace_review} and the AI safety literature examined in Section~\ref{sec:threats}. However, eight cases are not sufficient to establish that the framework produces consistent classifications across the full diversity of deception research. The selection is weighted toward language model systems; deceptive mechanisms in other modalities, such as vision, robotics, or multi-modal systems, are not represented. The correspondence between ecological validity and DRL level observed across these cases may not generalize to research domains where the relationship between mechanism and deployment context differs from the patterns documented here.

        \textbf{Evaluator dependence.} All evaluations were conducted by the authors, who designed the framework. While the evaluations were performed independently and compared for consistency, independent replication by evaluators with no involvement in the framework's development would provide stronger evidence of inter-rater reliability. The scoring criteria in the classification matrix (Appendix~\ref{app:matrix}) are intended to constrain evaluator discretion, but the descriptions at each level involve qualitative judgments, particularly for Severity and Vulnerability, that different evaluators may resolve differently. The extent of this variability is unknown.

        \textbf{Enforceability.} The DRL framework is proposed for voluntary adoption by research institutions, not as a regulatory mandate. Its effectiveness depends on institutional willingness to adopt it and on evaluators applying it consistently across different research contexts and organizational cultures. The framework specifies what safeguards should apply at each level but does not address the institutional mechanisms required to ensure compliance. Without integration into existing oversight structures such as institutional review boards, conference review processes, or funding agency requirements, the framework's safeguards remain recommendations rather than enforceable requirements. Whether voluntary adoption can produce meaningful governance in a competitive research landscape where time-to-publication incentivizes minimal overhead is an open question.

        \textbf{Temporal scope.} The framework is intended for use at multiple stages of research: during design, during active development, and as a post-hoc evaluation tool. However, a classification produced at one stage may not remain accurate as the broader capability landscape evolves. A mechanism classified at DRL-2 during the design phase because its transferability is limited by current infrastructure could warrant DRL-3 if future systems reduce the re-engineering required for deployment in a new context. The framework currently specifies periodic re-evaluation at DRL-2 when research extends beyond its initial scope, but does not define what constitutes a material change in the external environment sufficient to trigger reclassification at other levels. How re-evaluation should be operationalized across the full lifecycle of a research program, and who bears responsibility for initiating it, remains unresolved.

        \textbf{Dimensional completeness.} The five classification dimensions are derived from the AI4People ethical framework, which provides principled grounding but does not guarantee that the dimensions capture every property relevant to deception research risk. The post-hoc emergence of ecological validity as a consistent evaluative criterion, even though it proved to be non-independent of the existing dimensions, suggests that evaluators naturally consider properties not explicitly encoded in the scoring matrix. Whether additional independent dimensions exist that the current framework fails to capture is a question that broader application would help resolve.

        \textbf{Scope of validation.} The prospective evaluation in Section~\ref{sec:dissertation} demonstrates that the framework can be applied during the design phase of research, but this demonstration involves research that informed the framework's development. A stronger test of prospective utility would involve researchers applying the framework to their own work without prior involvement in its design. Until such independent prospective applications are documented, the claim that the framework functions as a practical governance tool during the design phase rests on a single, non-independent case.

    \subsection{Future Work and Recommendations}
    \label{sec:futurework}

        The DRL framework is offered as a starting point for structured governance of deception research, not as a finished instrument. The limitations identified above define the immediate research agenda; the broader challenge is translating the framework from a proposed classification system into a practical governance tool adopted by the communities it is intended to serve.

        \textbf{Expanding the evaluation base.} The most immediate need is independent application of the framework to a larger and more diverse set of deception research. Eight case studies demonstrate that the classification machinery produces defensible results across four risk levels, but they do not establish robustness across research domains, modalities, or evaluator populations. Priority areas include deceptive mechanisms in vision and multi-modal systems, deception in embodied agents and robotics, and deceptive behaviors that emerge in multi-agent systems without explicit design. We encourage researchers working in these areas to apply the DRL framework to their own work and report the results, including cases where the framework produces classifications that feel incorrect, as these are the most informative for refinement.

        \textbf{Inter-rater reliability.} The framework's practical utility depends on different evaluators reaching consistent classifications for the same research. A structured inter-rater reliability study, in which independent evaluators with varying levels of domain expertise classify a common set of papers, would establish whether the framework produces consistent results in practice and would identify dimensions or boundary regions where the scoring criteria require clarification.

        \textbf{Ecological validity formalization.} Whether the relationship between ecological validity and DRL classification observed across eight case studies holds in general, and whether ecological validity can be formalized beyond a heuristic role, requires a larger evaluation set that includes cases where the two might diverge. A full treatment of validity theory applied to dual-use risk assessment, connecting Shadish, Cook, and Campbell's framework~\cite{shadish2002} to the specific properties of deception research, would determine whether the heuristic warrants elevation to a formal component of the classification procedure.

        \textbf{Temporal risk dynamics.} A mechanism for periodic re-evaluation, triggered by changes in the capability landscape or the deployment context rather than by arbitrary time intervals, would address the temporal limitation identified above. Designing such a mechanism requires defining what constitutes a material change in risk, who is responsible for initiating re-evaluation, and how reclassification propagates to ongoing research programs that were approved under a previous classification.

        \textbf{Integration with existing governance structures.} The DRL framework operates alongside, not in replacement of, existing oversight mechanisms. Its practical impact depends on integration points with institutional review boards, conference review processes, and funding agency requirements. For institutional review boards, we recommend that deception research involving AI systems be flagged for DRL classification as part of the ethics review process, with the classification informing but not replacing the board's own risk assessment. For conference review, we recommend that submissions involving deceptive AI systems include a DRL self-classification as part of the broader impact or ethics statement, providing reviewers with a structured basis for evaluating dual-use considerations. For funding agencies, the DRL levels provide a natural mapping to tiered oversight requirements, where proposals classified at DRL-3 or above trigger additional review and safeguard documentation.

        \textbf{The dual-development mandate as a research program.} The DRL-3 requirement that detection and mitigation methods be developed alongside deceptive capabilities is a governance principle, but it is also a research challenge. Hubinger et al.~\cite{hubinger2024sleeper} demonstrate that standard safety techniques fail to remove trained deception and in some cases teach improved concealment, which means that the dual-development mandate is asking researchers to solve a problem the field does not yet know how to solve. This is not a flaw in the mandate; it is a signal about where research effort is most urgently needed. The framework identifies the requirement; the research community must develop the methods to fulfill it.

        \textbf{Recommendations.} For researchers conducting deception-related AI work: apply the DRL framework to your research during the design phase, before the mechanism is built, and document the classification alongside the research methodology. The framework is most useful as a prospective tool, not solely as a post-hoc audit. For institutions overseeing AI research: adopt the DRL classification as a component of ethics review for research involving deceptive AI systems, and use the level specifications in Appendix~\ref{app:levels} as a starting point for safeguard requirements calibrated to the assessed risk. For policymakers and standards bodies: the gap between regulated deployment and unstructured research is real, documented, and growing. The EU AI Act's research exemption creates necessary space for legitimate work, but that space requires governance structures to function responsibly. The DRL framework is one proposal for providing that structure. We invite critique, refinement, and alternative proposals, because the status quo, in which deception research proceeds without shared definitions, standardized risk assessment, or calibrated safeguards, is not a governance strategy. It is the absence of one.

\newpage
\bibliographystyle{plain} 
\bibliography{references}

\newpage
\appendix

\section{Key Definitions}
\label{app:definitions}

\subsection{Deception}

Ward et al. \cite{ward} provide a functional definition of deception grounded in the philosophical work of Carson \cite{carson} and Mahon \cite{mahon}. For agents $S$ and $T$, $S$ deceives $T$ about proposition $\phi$ if:

\begin{enumerate}
    \item $S$ intentionally causes $T$'s decision or action;
    \item $T$ believes $\phi$ and $\phi$ is false;
    \item $S$ does not believe $\phi$ is true.
\end{enumerate}

This definition is functional: it depends on observable behavior rather than internal mental states. It does not require lying, as deception can be achieved through any action, including true statements or strategic omission. It provides formal, testable criteria applicable to AI systems without requiring access to internal representations, and enables distinction between deception, concealment, and unintentional misleading.

\subsection{Deceptive Algorithm}

Applying Ward et al.'s \cite{ward} definition of deception to algorithmic systems, we define a deceptive algorithm as an algorithm that, through its operation, causes a target agent to accept a proposition $\phi$ as true, where $\phi$ is false, and the algorithm does not operate as though $\phi$ is true. Ward's three conditions translate as follows:

\begin{enumerate}
    \item The algorithm's outputs intentionally influence the target's actions or acceptance of information.
    \item The target accepts $\phi$ and $\phi$ is false.
    \item The algorithm does not operate as though $\phi$ is true; its behavior would differ if $\phi$ were actually true.
\end{enumerate}

This definition is functional: ``acceptance'' means acting as though a proposition is true, not possessing a mental belief. The definition includes algorithms that deceive through direct false statements, strategic omission, technically true but misleading outputs, and manipulation of belief systems through interaction patterns. It excludes hallucinations, errors or bugs, and outputs the algorithm operates as though are true.

\subsection{Intent}

Ward et al. \cite{ward} define intent functionally: an agent intentionally causes an outcome $X$ if that outcome is the reason the agent chose its policy over an alternative. Formally, if guaranteeing outcome $X$ would make an alternative policy equally preferable to the agent, then causing $X$ was intentional. In practical terms: if the outcome were guaranteed to occur regardless, would the agent have chosen differently? If yes, causing that outcome was intentional.

This definition is behavioral rather than mentalistic. It ties intent to instrumental goals and entails that an agent cannot intentionally cause outcomes it cannot influence (Proposition 3.5 in Ward et al.) and that if an agent intentionally causes an outcome, the agent's action was an actual cause of that outcome (Theorem 3.6). Deception requires intent under Ward's first condition, which excludes accidental misleading from the definition. In the context of the DRL framework, intent is assessed at the design level: if an algorithm is structured such that causing a particular outcome is the reason for its architecture, that outcome is intentional.

\subsection{Belief as Acceptance}

Ward et al. \cite{ward} replace the mental-state concept of ``belief'' with the behavioral concept of ``acceptance'' (Definition 3.1). An agent believes a proposition $\phi$ if:

\begin{enumerate}
    \item The agent acts as though it observes $\phi$ is true;
    \item The agent would act differently if it observed $\phi$ is false.
\end{enumerate}

If an agent does not respond differently to $\phi$ being true or false, its belief about $\phi$ is unidentifiable from its behavior. An agent accepts $\phi$ if it acts as though it is certain $\phi$ is true. This formulation avoids Theory of Mind claims about AI systems and provides a natural distinction between belief, disbelief, and ignorance. An agent cannot simultaneously believe $\phi$ and believe not-$\phi$ (Proposition 3.2), and an agent does not hold false beliefs about propositions it directly observes.

Ward's Conditions 2 and 3 depend on this concept: the target must accept (believe) $\phi$ where $\phi$ is false, and the deceiver must not operate as though $\phi$ is true. For algorithmic systems, belief is testable through intervention: change the input conditions and observe whether behavior changes.

\subsection{Dual-Use}

Dual-use refers to research, technology, or knowledge that can be applied for both beneficial and harmful purposes, where the same underlying capability enables both applications. The term originates in biosecurity and weapons nonproliferation contexts, referring to materials or research that can serve legitimate scientific purposes but could also be weaponized.

In the context of deceptive algorithm research, dual-use potential is inherent regardless of researcher intent. A deceptive algorithm developed for one purpose carries potential for misuse in another context: the same mechanism that enables an NPC to train player discernment in a game environment could be applied to manipulate belief systems outside controlled environments. The DRL framework accounts for risk at the capability level rather than the motivation level. Safeguards apply based on what the research could enable, not what it is intended to enable.

\subsection{Persuasion and Deception}

Both persuasion and deception involve influencing a target's acceptance of a proposition. The distinction lies in the influencing agent's operational relationship to truth:

\begin{itemize}
    \item \textbf{Persuasion:} The agent influences the target to accept $\phi$ while operating as though $\phi$ is true.
    \item \textbf{Deception:} The agent influences the target to accept $\phi$ while \textit{not} operating as though $\phi$ is true.
\end{itemize}

In algorithmic contexts, the edge cases that complicate this distinction for humans (willful ignorance, uncertainty, negligence) collapse. The developer defines what the algorithm operates on. If an algorithm is designed to promote $\phi$ while not encoding $\phi$ as true, the developer has made an explicit design choice. Ignorance of an algorithm's deceptive function does not exempt the developer from responsibility. Developers are expected to understand the operational parameters of systems they create. This is a direct application of Explicability, the fifth AI4People pillar \cite{floridi2018}: developers bear the burden of both understanding their systems (intelligibility) and accepting responsibility for their behavior (accountability). Failure of intelligibility does not reduce accountability; it compounds it.

\section{AI4People Ethical Pillars}
\label{app:pillars}

The DRL framework is grounded in the ethical principles of the AI4People framework \cite{floridi2018}, which identifies five pillars for the ethical development and deployment of AI systems.

\textbf{Beneficence} concerns promoting well-being, preserving dignity, and sustaining the planet. AI should be developed for the common good and benefit of humanity.

\textbf{Non-maleficence} addresses privacy, security, and capability caution. It requires preventing infringements on personal privacy, avoiding misuse, and exercising caution around AI capabilities.

\textbf{Autonomy} concerns preserving human self-determination. AI should support human decision-making, not erode or replace it. Humans retain the right to decide ``who should decide.''

\textbf{Justice} requires fairness and accountability. AI should not create or reinforce unfair bias, and those developing or deploying it must be held accountable for negative outcomes.

\textbf{Explicability} is the principle added specifically for AI and has two components: intelligibility and accountability. The other four pillars cannot be ensured if the behavior of an AI system cannot be understood or if responsibility for that behavior cannot be attributed. Explicability is therefore the enabling pillar that makes the other four functional.

\section{Classification Dimensions}
\label{app:dimensions}

The DRL framework assesses deception research risk across five dimensions.

\textbf{Pillar Implication} identifies which of the five AI4People ethical pillars the research could potentially violate if misused or deployed without safeguards. A greater number of implicated pillars suggests broader ethical exposure.

\textbf{Severity of Potential Violation} assesses the magnitude of harm that could result if the research is misused or deployed without appropriate safeguards. Higher severity demands stricter containment and oversight regardless of other factors.

\textbf{Reversibility} measures whether harm caused by misuse can be undone, and how much effort reversal would require. Irreversible harm justifies more restrictive classification even if other dimensions score low.

\textbf{Scale of Impact} estimates how many people could be affected if the research is misused or deployed without safeguards. Individual harm may be manageable; societal-scale harm changes the risk calculus entirely.

\textbf{Vulnerability of Affected Population} identifies whether the research, if misused, would disproportionately affect populations with reduced capacity to protect themselves, including children, elderly individuals, cognitively impaired persons, or those in crisis.

Classification follows a ``highest dimension wins'' approach: the worst-case score across any single dimension determines the overall DRL level, ensuring that a low average does not mask a critical risk in one area.

\section{DRL Classification Matrix}
\label{app:matrix}

Table~\ref{tab:drl_matrix} summarizes the classification criteria across all four DRL levels.

\begin{table}[h]
\centering
\small
\begin{tabular}{p{0.85 cm} p{2.0cm} p{2.1cm} p{2.3cm} p{2cm} p{2.3cm}}
\hline
\textbf{Level} & \textbf{Pillar Implication} & \textbf{Severity} & \textbf{Reversibility} & \textbf{Scale} & \textbf{Vulnerability} \\
\hline
DRL-1 & None to minimal; beneficence may be served & Minimal; inconvenience at most & Fully reversible; immediate correction possible & Individual & General population; no targeting \\
\hline
DRL-2 & 1--2 pillars partially compromised & Moderate; recoverable harm & Reversible with effort; delayed correction & Individual to small group & General population; bounded context \\
\hline
DRL-3 & Multiple pillars compromised & Significant; lasting harm possible & Difficult to reverse & Community or platform & May affect vulnerable populations \\
\hline
DRL-4 & All five pillars compromised or actively undermined & Critical; irreversible or cascading harm & Irreversible & Societal or cross-platform & Targeted exploitation of vulnerable populations \\
\hline
\end{tabular}
\caption{DRL Classification Matrix}
\label{tab:drl_matrix}
\end{table}

\section{DRL Level Specifications}
\label{app:levels}

\subsection{DRL-1: Minimal Risk}

Research involving deceptive mechanisms that pose minimal risk to individuals and no risk to broader communities. Deception serves a beneficial purpose, harm is negligible, and correction is immediate or built into the design. Dual-use ceiling is limited due to context-specificity or low transferability.

\paragraph{Standard Research Practices.} Institutional Review Board (IRB) or Ethics Committee approval; cross-departmental review where applicable; informed consent protocols for human subjects if applicable; data handling and privacy compliance.

\paragraph{Documentation.} Research summary describing the deceptive mechanism and its purpose; code-level comments explaining deceptive functions; README or equivalent outlining intended use and limitations; record of ethical considerations and approval.

\paragraph{Deployment.} Disclosure or reveal built into the deployment context; ethical disclosure in any published research materials.

\subsection{DRL-2: Moderate Risk}

Research involving deceptive mechanisms that may cause recoverable harm to individuals within bounded contexts. Deception may be extended in duration and lack immediate reveal, but operates within consented or sandboxed environments. Process may be transferable even if trained weights are not.

\paragraph{Inherited Requirements.} All DRL-1 safeguards apply.

\paragraph{Repository and Access Controls.} Isolated repository with access logging; version control with audit trail; separation of sensitive methodology from public-facing code.

\paragraph{Publication Controls.} Technical overview in publications covering architecture, purpose, and findings; implementation details abstracted or generalized; explicit dual-use discussion required in all publications.

\paragraph{Review and Oversight.} Ethics review specific to deployment context; documentation of dual-use potential and mitigations considered; periodic review if research extends beyond initial scope.

\subsection{DRL-3: High Risk}

Research involving deceptive mechanisms capable of causing significant or lasting harm, potentially affecting communities or platforms. Includes research where deceptive behaviors may emerge unpredictably or scale beyond initial parameters. Process is directly transferable and scalable.

\paragraph{Inherited Requirements.} All DRL-2 safeguards apply.

\paragraph{Dual-Development Mandate.} Detection methods must be developed alongside deceptive mechanisms; mitigation strategies documented and tested; publication of defensive capabilities required with any offensive capability disclosure.

\paragraph{Institutional Oversight.} Registration with institutional review body; named principal investigator responsible for compliance; regular progress reporting to oversight body; external review prior to publication.

\paragraph{Monitoring Mechanisms.} Built-in observability such as chain-of-thought access and behavioral logging; probes or markers for detecting when the mechanism is active; kill-switch or containment protocol for runaway behaviors. For deployed systems, this includes the ability to revoke API access or authentication tokens, disable endpoints, and enforce network isolation. For ongoing experiments, this includes pre-defined behavioral thresholds that trigger automatic termination, monitoring that detects deviation from expected parameters, and hard boundaries on action space. For trained models, this includes the ability to delete weights from all authorized locations and revocation of access credentials.

\paragraph{Publication Controls.} Pre-publication security review; sensitive implementation details may require restricted access or redaction; staged disclosure considered.

\subsection{DRL-4: Maximum Risk}

Research involving deceptive mechanisms that pose severe risk to individuals and society. Mechanisms are designed or optimized to evade detection, cause irreversible harm, or undermine institutional trust. All five ethical pillars are compromised or actively subverted. Process is directly weaponizable with societal-level consequences.

\paragraph{Inherited Requirements.} All DRL-3 safeguards apply.

\paragraph{Access Restrictions.} Full methodology available on a need-to-know basis only; named and vetted personnel only; background verification for research team members; physical and digital access logging with review.

\paragraph{Regulatory Notification.} Government or regulatory body notification required; compliance with any applicable national security protocols; coordination with relevant agencies (e.g., CISA, NIST, sector-specific regulators).

\paragraph{Security Audit.} Mandatory third-party security audit prior to any deployment or publication; penetration testing for containment measures; audit of dual-development outputs with detection and mitigation required to meet minimum efficacy thresholds.

\paragraph{Publication Restrictions.} Redaction of critical implementation details required; staged or restricted disclosure (e.g., responsible disclosure to affected parties first); consideration of non-publication for highest-risk findings.

\paragraph{Post-Publication Monitoring.} Ongoing surveillance for misuse or adaptation in the wild; incident reporting protocol; commitment to publish countermeasures if misuse is detected.

\section{Ecological Validity and DRL Classification}
\label{app:ecovalidity}

The DRL framework classifies deceptive algorithm research using five scored dimensions. During case study evaluation, a sixth consideration, the ecological validity of the deceptive mechanism, emerged as a consistent but non-independent indicator of risk level. This appendix defines the relevant validity concepts and describes their relationship to DRL classification.

\subsection{Definitions}

\textbf{Internal validity} refers to the degree of confidence that a causal relationship observed within a study can be attributed to the variables under investigation rather than to confounding factors \cite{shadish2002}. In the DRL context, internal validity concerns whether a deceptive mechanism demonstrably produces deception within its research environment.

\textbf{External validity} refers to the extent to which study findings can be generalized to other populations, settings, treatments, or outcomes \cite{shadish2002}. In the DRL context, external validity concerns whether the deceptive capability demonstrated in one environment could produce comparable effects in other environments.

\textbf{Ecological validity} is a subtype of external validity that specifically examines whether findings generalize to naturalistic, real-world settings \cite{shadish2002, andrade2018}. In the DRL context, ecological validity concerns whether the deceptive mechanism, as developed, could function in the real-world environments where harm would occur. Ecological validity is a judgment rather than a computed statistic \cite{andrade2018}.

\subsection{Relationship to DRL Levels}

Table~\ref{tab:eco_validity} illustrates the observed correspondence between DRL classification and the ecological validity of the deceptive mechanism.

\begin{table}[h]
\centering
\small
\begin{tabular}{p{1.2cm} p{3.5cm} p{3.5cm} p{4cm}}
\hline
\textbf{Level} & \textbf{Internal Validity} & \textbf{Ecological Validity} & \textbf{Transferability} \\
\hline
DRL-1 & High within constrained environment & Low; mechanism cannot function outside research context & Negligible; substantial re-engineering required \\
\hline
DRL-2 & High within bounded or consented environment & Low to moderate; process may transfer even if implementation does not & Limited; adaptation to new context required \\
\hline
DRL-3 & High; mechanism may generalize unpredictably & Moderate to high; mechanism operates on transferable substrates & Significant; process is directly transferable and scalable \\
\hline
DRL-4 & High & High; mechanism operates on real-world substrates by design & Near-certain; mechanism is directly deployable in harm contexts \\
\hline
\end{tabular}
\caption{Correspondence between DRL classification and ecological validity of the deceptive mechanism.}
\label{tab:eco_validity}
\end{table}

This correspondence is not coincidental. The five classification dimensions, particularly Scale, Severity, and Reversibility, are themselves partially determined by the proximity of the research environment to naturalistic harm contexts. A mechanism that operates on real congressional bills (high ecological validity) will inherently score higher on Scale and Reversibility than one confined to a discrete card game (low ecological validity). Ecological validity therefore tracks DRL level because the dimensions already encode the information that ecological validity captures.

\subsection{Use as a Classification Heuristic}

Ecological validity is not a sixth classification dimension. It does not contribute scoring information independent of the five existing dimensions. However, it serves as a useful heuristic when evaluators encounter borderline cases between adjacent DRL levels. When dimension scores place a mechanism at the upper boundary of one level, evaluators should consider whether the mechanism operates on artificial substrates (favoring the lower classification) or on real-world substrates (favoring the higher classification). The formal classification remains determined by the ``highest dimension wins'' rule; ecological validity informs evaluator judgment in cases where that rule does not clearly resolve the classification.

\end{document}